\documentclass[11pt,twoside]{article}

\usepackage{asp2006}
\usepackage{epsf}
\usepackage{psfig}
\usepackage{lscape}

\begin{document}
\title{On the deduction of galactic abundances with evolutionary neural networks}  
\author{M. Taylor and A. I. D\'iaz}  
\affil{Grupo de Astrof\'isica, Departamento de F\'isica Te\'orica,\\
Universidad Aut\'onoma de Madrid (UAM), 28049 Madrid, Spain}    

\begin{abstract}
A growing number of indicators are now being used with some confidence to measure the metallicity($Z$) of photoionisation regions in planetary nebulae, galactic HII regions(GHIIRs), extra-galactic HII regions(EGHIIRs) and HII galaxies(HIIGs). However, a universal indicator valid also at high metallicities has yet to be found. Here, we report on a new artificial intelligence-based approach to determine metallicity indicators that shows promise for the provision of improved empirical fits. The method hinges on the application of an evolutionary neural network to observational emission line data. The network's DNA, encoded in its architecture, weights and neuron transfer functions, is evolved using a genetic algorithm. Furthermore, selection, operating on a set of 10 distinct neuron transfer functions, means that the empirical relation encoded in the network solution architecture is in functional rather than numerical form.  Thus the network solutions provide \textit{an equation} for the metallicity in terms of line ratios without \textit{a priori} assumptions. Tapping into the mathematical power offered by this approach, we applied the network to detailed observations of both nebula and auroral emission lines from $0.33\mu m-1\mu m$ for a sample of 96 HII-type regions and we were able to obtain an empirical relation between $Z$ and $S_{23}$ with a dispersion of only $0.16$ dex. We show how the method can be used to identify new diagnostics as well as the nonlinear relationship supposed to exist between the metallicity $Z$, ionisation parameter $U$ and effective (or equivalent) temperature $T*$.
\end{abstract}

\section{Introduction}
Emission lines due to photoionisation of nebulae by massive stars are the most powerful indicators of the chemical evolution of galaxies in both the near and the intermediate redshift universe. The wealth of new data coming from galaxy spectroscopic surveys and from observations with integral field units of nearby galaxies allow for a detailed inventory of the chemical composition of star forming galaxies. Abundance determinations in HII regions are relatively straightforward to calculate if the electron temperature, $T_{e}$, can be measured directly from the observations. However, in many cases, the $T_{e}$ diagnostic lines are too faint to be detected due to the strong cooling effect of metals in these regions. Faced with this obstacle, \citet{1979Pagel} and \citet{1979Alloin} pioneered methods based on the $O_{23}$ parameter that allow one to estimate the metallicity $Z$ \footnote {Throughout the paper "metallicity" is used with the meaning of "oxygen abundance"} using strong lines only.
\newline
The strong line methods assume that all HII regions are essentially characterised by their metallicity. However, the metallicity $Z$ and other empirical parameters such as the hardness of the ionising radiation (which can be inferred from the spectral energy distribution and parameterised through the effective or equivalent temperature $T*$) and the ionisation parameter $U$ are nonlinearly inter-linked (\cite{1988Vilchez} and \cite{1991Diaz}) but the exact nonlinear relation between $Z$, $U$ and $T*$ is still unknown. Observed line ratio diagnostics are central to the whole endeavour, hence the large amount of literature devoted to the identification of abundance indicators and their calibration (\cite{2005Perez} and \cite{2004Stasinska}). In fact, these are two separate and important issues - the choice of indicator, and the proper abundance calibration. Although a plethora of different indicators have been proposed, such as $O_{23}$, $O_{3}N_{2}$, $N_{2}$, $S_{23}$, $Ar_{3}O_{3}$ and $S_{3}O_{3}$, their full calibration over the total range of metallicity has not yet been possible with confidence due to a lack of data and understanding of the physics in metal-rich HII regions (\cite{2006Bresolin} and \cite{2006Pilyugin}). Temperature fluctuations are known to play an important role (\cite{1967Peimbert} and \cite{2006Peimbert}) and the differences between abundances calculated using optical recombination and collisionally-excited lines (that can vary by up to a factor of 70 (\cite{2006Liu}) appear to be related to atomic physics in the infrared. With this in mind we have developed a new artificial intelligence technique to find, from the optical emission lines, the best indicators so that the method can be applied with confidence to forthcoming infrared emission line data.
\newline
Here, we present our results for a study of HII-like regions using the $S_{23}$ diagnostic to illustrate the basic principles. We have selected $S_{23}$ from the most commonly-used indicators as it has low dispersion and is quasi-linear up to and slightly above solar metallicity (see table 4 in \cite{2005Perez}). Of course, at high metallicities it is expected to turn over, following a parabolic and double-branched curve as for $O_{23}$ due to effective cooling from $S$ ions. Our approach to the problem uses a genetic network to seek out the empirical relation between metallicity and $S_{23}$.

\section{SAGAN}
The \textit{Scale-invariant And Genetically-Adapted neural Network} (SAGAN) has been developed to deduce empirical laws from observational data (\cite{2006TDW}). It uses a simple backpropagation error rule \textbf{and is capable of correctly identifying an equation} for the functional relationship $f$ between the independent variable $x_{n}$ and the dependent physical variables  $(x_{1},x_{2},\cdots,x_{n-1})$ associated with training data. Contrary to the common practice of subjectively choosing from a library of fitting functions, the network, objectively and unambiguously finds the function that best fits both the training and evaluation data, and hence has great generalisation potential. Even when fed with multi-dimensional data having mesurement errors of 10\%, the network successfully identifies exact equations (\cite{2006TDW}). Key to its success is a genetic algorithm that evolves the network DNA encoded in the distribution of neuron weights, connectivity and transfer functions. The neuron transfer functions undergo selection from a list of 10 mathematical functionals and are what provide semantic structure to the network solutions found.
\newline
Another feature central to the success of SAGAN, is that it is a dimensionless network; meaning that it benefits from the dimensionality reduction associated with applying the $\pi$-theorem (\cite{1914Buckingham}) to the data (\cite{2006TD}). In addition to producing a network that is topologically simpler, the degeneracy inherent in the dimensionless groups $\pi_{j}$ facilitates learning and leads to an error reduction of several orders of magnitude (\cite{2006TDW}). Metallicity and line ratios are, by definition, dimensionless quantities; nevertheless SAGAN exploits the degeneracy associated with combinations of dimensionless groups of variables to expand its search space automatically. SAGAN operates as a supervised neural network. The metallicity $Z$ is calculated independently using the auroral line method (incorporating $T_{e}$ and ionisation correction factor effects) and provides the supervising component. SAGAN then cycles through a genetic algorithm, evolving its network DNA. In each generation, a population ($p$) of 200 "parent" network solutions is generated using the standard genetic operators of mutation, replacement and cross-over. Those parents providing the best fit to the data are kept, replicated and bred through cross-over to generate the basis for mutating the next generation. The best individuals in each generation are retained so that they are not lost from the gene pool. The process continues until the network error $E$,

\begin{equation}
	E=\frac{1}{p}\sum_{p}\left[f(x_{1},\cdots,(x_{n-1})-(x_{n})_{p}\right]^{2}
\end{equation}

converges to a minimum value or until the maximum number of generations specified is reached. The correct empirical law is found when the encoded network DNA has evolved to the point that it is capable of decoding the causal relation existing between input and output observables. In the case of the relation between $Z$ and $S_{23}$, the network error is equal to the dispersion in dex. The minimum error solution is then converted into an empirical equation by reading from right to left and summing nodes.

\section{Results}

\begin{figure}[h]
\begin{center}
\vspace{8cm}
 \includegraphics{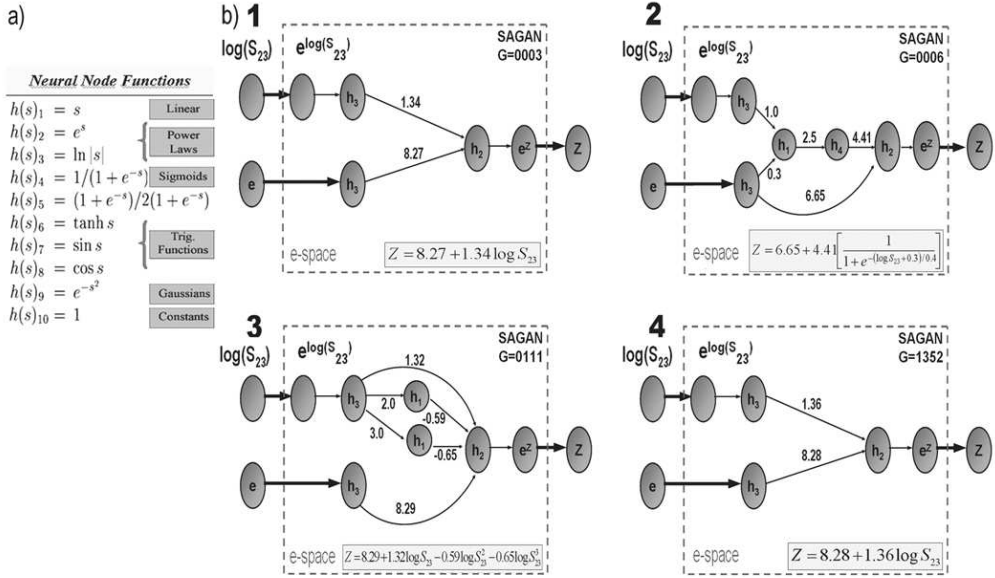}
\end{center}
\caption{a) The list of neural transfer functions and b) the empirical networks ($1-4$) deduced by SAGAN during its evolution. Note the transformation to exponential space (e-space) needed to handle negative values of $log(S_{23})$ while keeping the network dimensionless.}\protect\label{fig1}
\end{figure}

We collected optical emission line data for 121 HII-type photoionisation regions that included the sample of 108 HIIGs, GHIIRs and EGHIIRs studied by \cite{2000Diaz} plus an additional 8 HIIGs from \cite{2006Hagele} and 5 metal-rich GHIIRs taken from \cite{2004Bresolin}. Of the 121 objects, we selected 96 for which both de-reddened nebular and auroral lines of $S$, $O$ and $N$ in the optical existed, along with an accurate measurement of the metallicity $Z$ complete with measurement errors. The value of the $S_{23}$ parameter (\cite{2000Diaz}),

\begin{equation}
	S_{23}=\frac{[S_{II}]\lambda\lambda6717,6731+[S_{III}]\lambda\lambda9069,9532}{H_{\beta}}
\end{equation}

for each object, along with its metallicity, provided the training data set for SAGAN. In order to create a validation data set, we generated a further 96 points within the error bars of the training data using a random number generator based on the normal distribution. During the evolution process, SAGAN outputs the network solution for those generations during which there is a drop in the dispersion (network error). Figure \ref{fig1} shows the network solutions associated with the largest error drops. Although some 1352 generations taking 2 days of runtime on a dual 3Ghz PC were required, SAGAN achieved a final linear fit having a dispersion of only $0.16$ dex.

In order to obtain the empirical law, one has to read from right to left from the output variable back through the network to the input variables, applying the neuron transfer functions and summing the weighted node paths. The empirical relation for figure \ref{fig1}$b)4$ for example is,

\begin{equation}
	e^{Z}=e^{\left\{1.36ln\left(e^{log\left(S_{23}\right)}\right)\right\}+8.28ln(e)}.
\end{equation}

Noting that $ln(e^{x})=x$ and equating the exponents of $e$, we obtain,

\begin{equation}
	Z=8.28+1.36log\left(S_{23}\right).
\end{equation}

\begin{figure}[h]
\begin{center}
\vspace{11.5cm}
 \includegraphics{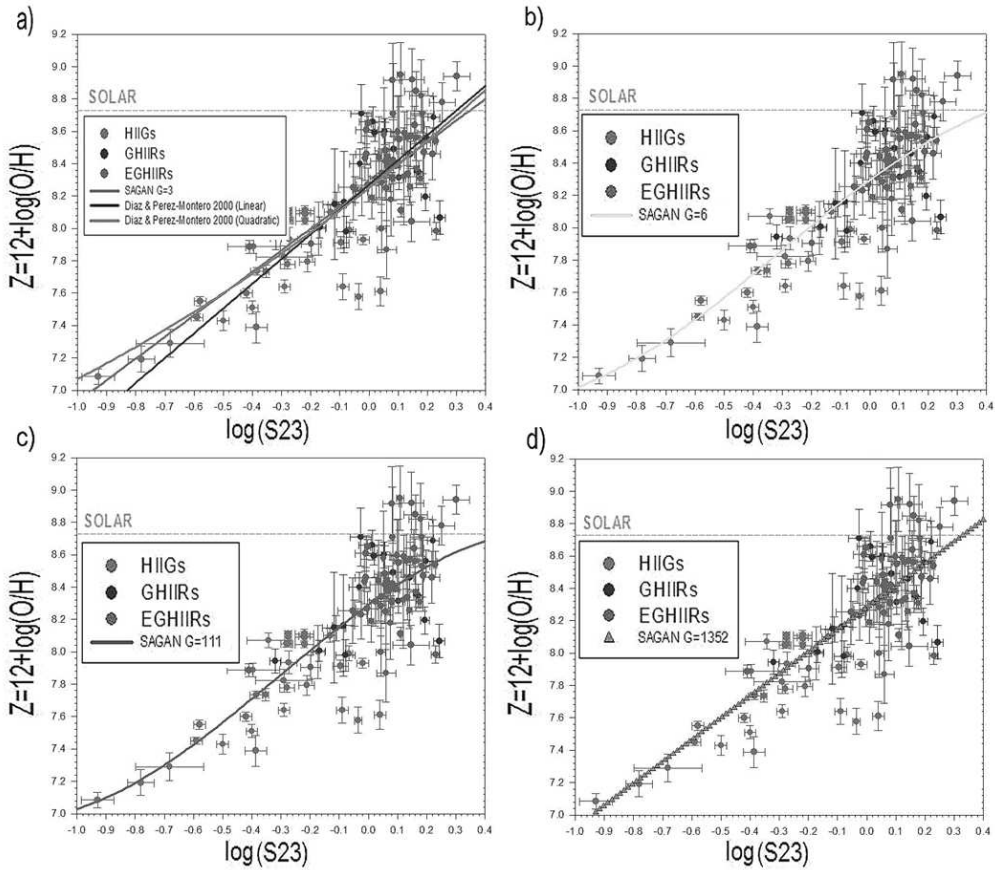}
\end{center}
\caption{The empirical fits obtained by SAGAN during generations 3, 6, 111 and 1352 of its evolution. In $a)$ we also plot the linear and quadratic fits obtained by \cite{2000Diaz}.}\protect\label{fig2}
\end{figure}

The biggest improvements can be seen to be associated with topological changes in the network. In figure \ref{fig2}, we have extracted the empirical laws from the network and plotted them over the training data. In figure \ref{fig2}$a)$, we can see that after only 3 generations, SAGAN identified a linear fit to the data having a dispersion of $0.22$ dex. Although this is slightly higher than the $0.20$ dex dispersion published for $S_{23}$ (\cite{2000Diaz}), SAGAN rapidly improved upon this in its sixth generation shown in figure \ref{fig2}$b)$ with a sigmoidal fit to the data having a dispersion of $0.183$ dex. The next improvement occured in the 111th generation in figure \ref{fig2}$c)$ when SAGAN found a cubic fit to the data having a dispersion of $0.181$ dex. After 2 days runtime, and after 1352 generations, SAGAN unexpectedly found in figure \ref{fig2}$d)$ a linear fit with a dispersion of only $0.16$ dex. We left SAGAN running until its 8500th generation and no further improvement was found. The return to a linear fit after so many generations of evolution needs some comment. Why did SAGAN not find this linear fit directly in generation 3? It appears that it has investigated other functional forms and then after a process of significant learning has adjusted the gradient of the linear fit to minimise the dispersion. Contrary to least squares fitting routines, SAGAN, has experimented with numerous fitting functions and then adjusted the most suitable one. We consider this the main advantage of our network - that it objectively seeks the best fitting function.

\section{Conclusions and Future Plans}
We have shown that SAGAN produces accurate results that are robust to errors on the training data. Furthermore, the symbolic equations extracted are the best empirical laws to date that fit observational data of HII regions. The fits are consistent with earlier results obtained by subjectively fitting the data with linear and quadratic least-square fitting routines (\cite{2000Diaz}).
\newline
With the sample of 96 objects for which we have auroral and nebular emission lines of $O$, $N$ and $S$, we are now investigating $n$-dimensional indicators of the form $Z=f(x_{1},x_{2},\cdots,x_{n})+c$ with each $x_{i}$ involving a line ratio and $c$ is a constant. We will shortly report on SAGAN's findings for these potentially new indicators. Additionally, we are compiling a list of relations involving emission line ratios of $O$, $N$ and $S$ that are correlated with metallicity $Z$, ionisation parameter $U$ and effective or equivalent temperature $T*$ in order to deduce the nonlinear relationship $f(Z,U,T*)=0$ for application to infrared data in the near future.

\acknowledgements
MT would like to thank the Universidad Aut{\'o}noma de Madrid Astrophysics Group for their hospitality and financial support. Part of this work was funded by the project "Estallidos de Formaci{\'o}n Estelar en Galaxias" (AYA2001-3939-C03) from the Spanish Ministry of Science, and the Communidad de Madrid ASTROCAM project (S-0505/ESP/000237). We would also like to thank Johann Werner (University of St\"uttgart) for the providing the original code upon which SAGAN is based.

\end{document}